\documentclass[fleqn,usenatbib]{mnras}

\usepackage{newtxtext,newtxmath}

\usepackage[T1]{fontenc}

\DeclareRobustCommand{\VAN}[3]{#2}
\let\VANthebibliography\thebibliography
\def\thebibliography{\DeclareRobustCommand{\VAN}[3]{##3}\VANthebibliography}

\usepackage{graphicx}	
\usepackage{amsmath}	

\title[Limits of coherent curvature radiation]{Coherent curvature radiation: maximum luminosity and high-energy emission}

\author[A. J. Cooper et al.]{
A. J. Cooper,$^{1,2}$\thanks{E-mail: a.j.cooper@uva.nl}
R. A. M. J. Wijers,$^{1}$
\\
$^{1}$API, Anton Pannekoek Institute for Astronomy, University of Amsterdam, Science Park 904, 1098 XH Amsterdam, the Netherlands \\
$^{2}$ASTRON, Netherlands Institute for Radio Astronomy, Oude Hoogeveensedijk 4, 7991 PD, Dwingeloo, the Netherlands
}

\date{Accepted XXX. Received YYY; in original form ZZZ}

\pubyear{2015}

\begin{document}
\label{firstpage}
\pagerange{\pageref{firstpage}--\pageref{lastpage}}
\maketitle

\begin{abstract}
High brightness temperature radio transients such as pulsars and fast radio bursts require the coherent radiation of particles. The antenna class of coherent radiation models require a large number of charged particles radiating in phase, therefore the particles must be spatially confined and have well-aligned velocities. Given these necessary conditions, we look at the magnetic field induced by the currents associated with coherently emitting accelerated particles and consider the interaction between the radiating particles and the induced magnetic field. We find a maximum luminosity of coherent curvature radiation that depends on source parameters such as surface magnetic field and neutron star spin period. We find that coherent radio emission across all luminosities can be explained by coherent curvature radiation and suggest it could be universally responsible for both FRBs and extreme galactic sources. Using the Crab Pulsar as an example, we constrain the emission parameters and origin of the most extreme nanoshots to within 60km of the pulsar surface assuming coherent curvature radiation. In agreement with recent observations, we also predict simultaneous X-ray emission from small-scale particle gyration due to the induced field.

\end{abstract}




\section{Introduction}
\label{Introduction}
Coherent radiation is required for luminous, short duration radio transients, where the high brightness temperature cannot be explained by relativistic beaming alone \citep{Pietka2015,Melrose2017}. The two primary examples of these very high brightness temperature astrophysical sources are pulsars and Fast Radio Bursts (FRBs). Coherent emission can be broadly classified into either maser emission or antenna emission which requires spatial clustering of particles \citep{Zhang2020}. Coherent curvature radiation is a model of the latter and here we examine limits of this radiation model.
\par
Coherent curvature radiation has been used to explain high brightness temperature emission from pulsars \citep{Sturrok1971,RudermanSutherland1975, Yang2018}. More recently, this model has become one of the front-running radiation models of FRBs, where the conditions for the coherent emission of a large number of particles are found in the inner magnetospheres of highly-magnetized neutron stars known as magnetars \citep{Katz2016,Cordes2016,Kumar2017,Ghisellini2018}. In general, these models suggest that acceleration gaps of unscreened electric field parallel to the magnetic field lines, $E_{\parallel}$, accelerate particles along magnetic field lines producing curvature radiation. However, there are many open questions in terms of how these particles radiate in phase \citep{Lyubarsky2021}; for example what are the sufficient conditions for particles to act coherently in this manner. In this Letter we focus on two basic, necessary properties of coherent radiation and look at the electrodynamic interactions between radiating particles. 
\par
Firstly, we use the fact that coherently emitting particles must do so from a region no bigger than a comoving size $R_{\rm coh} < \gamma \lambda$ where $\lambda$ is the wavelength of observed emission. Secondly, we assume that the particles' velocities must not be misaligned by more than a factor of $\gamma^{-1}$, i.e. $\frac{\delta \textit{\textbf{p}}}{|p|} < 1/\gamma$. This limit is used in the coherent curvature radiation model of \cite{Kumar2017}, where authors suggest the induced perpendicular field due to the current of accelerated particles $B_{\perp}$ must be smaller than the local field $B$ along which the electrons stream by a factor of $\gamma$:
\begin{equation}
    B \geq \gamma B_{\perp}
    \label{eq:B_constraint}
\end{equation}
For magnetic fields approaching $B_c = \frac{m_{\rm e}^2 c^3}{e \hbar} = 4.4 \times 10^{13}$ G, the excitation energy of the first electron Landau level becomes comparable to the electron rest mass. In \cite{Kumar2017}, the authors suggest that the local magnetic field in which bright FRBs radiate must be $\gtrsim 10^{14}$G such that particles are not dislodged from the ground state despite perturbation, and coherence is maintained. 
\par
In Section \ref{sect:Induced magnetic field} we look at the magnetic field induced by accelerated particles and re-derive Eq. \ref{eq:B_constraint} by considering the perturbation of particles' momenta. In Section \ref{sect:Limitations} we look at the constraints due to this perturbation, and find an upper limit of the lorentz factor of the coherently radiating particles. In Section \ref{sect:Predictions} we find an upper limit for radio luminosity of coherent curvature radiation for sub-critical magnetic fields, in agreement with the luminosity gap between extreme galactic pulsar emission \citep{Hankins2003,Kuiack2020} and extra-galactic FRBs \citep{Petroff2019,Zhang2020}. We apply these constraints to the giant pulses observed from the Crab Pulsar and constrain the emission to within 60km of the NS surface. Furthermore, the most extreme crab nanoshots \citep{Hankins2007} must originate on the surface of the star if produced by coherent curvature emission. We also look at high-energy emission due to the small-scale perturbations of particles' motion, and suggest this could explain the recent detection of enhanced X-ray activity emission associated with giant radio pulses \citep{Enoto2021}. We conclude with a short discussion in Section \ref{sect:Conclusion}. We use convenient notation $X_{\rm n} \equiv X/10^n$ throughout.

\section{Induced magnetic field}
\label{sect:Induced magnetic field}
Consider a bunch of electrons or positrons that are spontaneously accelerated along curved magnetic field lines $B$ by an strong electric field parallel to the magnetic field lines $E_{\parallel}$, where $E < B$. The origin of the accelerating electric field or charge creation event is not discussed here, but could be for example a magnetic reconnection event. The acceleration length scale $l_{\rm acc} = \gamma m_{\rm e} c^2/q E_{\parallel} \approx 10^{-2}\,{\rm cm}\; \gamma_{3} \, E_{\parallel,8}^{-1}$ is assumed to be smaller than the spatial scale of the radiation patch throughout. To observe coherent radiation at a wavelength $\lambda$ the particles must at least obey the following conditions:
\begin{equation}
 R_{\rm coh} \leq \gamma \lambda = \gamma c \nu^{-1} 
\label{eq:R_coh}
\end{equation}
\begin{equation}
    \frac{\delta \textit{\textbf{p}}}{|p|} < \frac{1}{\gamma}
\label{eq:deltamomenta}
\end{equation}
These two equations tell us that the particles' positions and momenta respectively must be well confined in order to radiate coherently. Eq. \ref{eq:R_coh} applies to the source's longitudinal extent, but the transverse coherent region can be larger than $\gamma \lambda$ by a factor $\eta^{1/2}$ due to photon arrival delay which depends on the distance to the source's trigger $l_t$ \citep{Kumar2017}. We take $\eta = 1$ for simplicity and because our results depend very weakly on this parameter, such that the total comoving volume is $V^{'} = \eta \gamma^3 \lambda^3$. We further assume that the bunch has propagated a distance $l_t$ from the trigger such that coherent region is transversely causally connected: $R_{\rm coh} < \frac{l_t}{\gamma^2}$, or $l_t > \gamma^3 \lambda$. Multiple longitudinal patches of coherent radiation $N_{\rm p,l}$ may be consecutively observed as discussed in Section \ref{sect:duration}.
\par
Particles streaming along the guiding magnetic field line $B$ induce a current which in turn induces a magnetic field. This secondary field $B_{\perp}$ can perturb the particles, limiting coherent emission. We consider a bunch of $N$ particles confined in a space with a co-moving radius $R_{\rm coh}$ moving at $\gamma$, where the $E_{\parallel}$ acceleration balances radiation losses. These particles produce a current density J such that:

\begin{equation}
J = 2 n_{\rm e} c e  = 2 \gamma n_{\rm e}^{'} c e 
\label{eq:current}
\end{equation}
The co-moving electron density $n_{\rm e}{'} = n_{\rm e}/\gamma$ where $n_{\rm e}$ is in the lab frame, and we have assumed that the particles are accelerated to approximately $v=c$. Assuming this current is steady on short time scales, it induces a magnetic field perpendicular to the current \citep{Kumar2017}:
\begin{equation}
    \nabla \times B \approx \frac{B_{\perp}}{R_{\rm coh}} = \dfrac{4 \pi J}{c} 
\end{equation}
\begin{equation}
\begin{split}
        B_{\perp} &= \dfrac{4 \pi R J}{c} = 8 \pi R_{\rm coh} e n_{\rm e} = 8 \pi e n_{\rm e} \gamma c \nu^{-1}\\
    \label{eq:B_perp}
\end{split}
\end{equation}
\subsection{Particle motion}
\label{sect:particle_motion}
Consider the motion of these particles due to the induced field. We define $B = B_{z}$, $B_{\perp} = B_{\perp,\phi}$, $E_{\parallel} = E_{\parallel,z}$, such that the z-axis is locally tangent to the curved dipole magnetic field lines. Particles follow the total field line $B_z + B_{\perp}$ resulting in helical motion about $B_z$ with a pitch angle $\alpha = v_{\phi}/c =  B_{\perp}/B_z$, where we have assumed $v_z \approx c$ due to the strong $E_{\parallel}$. Given this, we can see how the momenta condition in Eq. \ref{eq:deltamomenta} is the same as the condition in Eq. \ref{eq:B_constraint}. The particle acceleration and gyroradius are:

\begin{equation}
\begin{split}
a_r = \frac{v_{\phi}^2}{r} &=  - \frac{q v_{\phi} B_{z}}{\gamma m_{\rm e}}  -  \frac{q v_{z} B_{\perp}}{\gamma m_{\rm e}} =  - \frac{2 q c B_{\perp}}{\gamma m_{\rm e}}
\end{split}
\end{equation}

\begin{equation}
    r_g = r = \frac{\gamma m_{\rm e} c B_{\perp}}{2 q B_z^2} = \frac{\gamma m_{\rm e} c \alpha}{2 q B_z} = \frac{\gamma m_{\rm e} v_{\phi}}{2 q B_z}
    \label{eq:gyroradius}
\end{equation}
In Section \ref{sect:highenergy} we will suggest this particle acceleration along field lines, which is the equivalent motion as synchrotron gyration about $B_z$, results in high-energy radiation. It is possible for accelerated particles to emit coherently for a short period of time $t < \delta t$ before the force due to the induced field $B_{\perp}$ has imparted sufficient momentum to destroy coherence, however this timescale is extremely short: $dt < \frac{2 \pi r_g}{v_{\phi}} = 10^{-26} \; {\rm s} \: \gamma_{3} \, B_{11}^{-1}$.

\section{Limits of coherent curvature radiation}
\label{sect:Limitations}

\subsection{Constraint due to spatial confinement and absorption}
For coherent radiation we require that particles are spatially confined via Eq. \ref{eq:R_coh}, therefore we should also require that $r_g < R_{\rm coh}$:
\begin{equation}
\begin{split}
    \frac{\gamma m_{\rm e} c B_{\perp}}{2 q B^2} &<  \frac{\gamma c}{\nu} \\
    n_{\rm e} \gamma &< \frac{B^2}{4 \pi m_{\rm e} c } \approx 2 \times 10^{37} \: B_{11}^2
\end{split}
\end{equation}
Where we have used the lab frame number density and Eq. \ref{eq:R_coh}, and $B$ is the local magnetic field strength of $B_z$. We find that the gyration radius $r_g$ is small compared to the coherent emission radius $R_{\rm coh}$, and therefore this does not meaningfully constrain the emission. In fact, the particle gyroradius derived in Eq. \ref{eq:gyroradius} could help explain why coherently emitting particles can stay confined spatially for the duration of emission despite electrostatic repulsion. The coherent curvature radiation will have a X-mode component transverse to both the local magnetic field $\vec{B}$ and the wave-vector $\vec{k}$ \citep{Kumar2017}. This component easily escapes even high particle density sources as it may propagate in a magnetized plasma if: $\omega > \omega_p^2 / \omega_B > \; 10^{-11}  \:  B_{11}^{-1} \, n_{{\rm e},12} \, \gamma_{3}^{-1} \, {\rm Hz}$ when $\omega_B > \omega_p$ \citep{AronsBarnard1986}.

\subsection{Constraint due to particle gyration cooling}
\label{sect:gyrationcooling}
The particles follow the total $B_z + B_{\perp}$ field lines along a helical path with pitch angle $\alpha = B_{\perp}/B_z$. The particles' path is identical to synchro-curvature radiation \citep{Cheng1996,Kelner2015} despite following the total field line, and this gyration leads to additional incoherent cooling. For particles in the coherent region, we must compare incoherent synchrotron radiation due to gyration to the large scale coherent curvature radiation to find the dominant cooling mechanism:
\begin{equation}
\begin{split}
    P_{\rm sync} &> P_{\rm curv} \\
    \frac{1}{4} N \pi c \sigma_T B^2 \gamma^2 \alpha^2 &> \frac{2 (N e)^2 c \gamma^{4}}{3 \rho^{2}} \\
    \gamma &< \bigg(\frac{24 \rho^2 \pi \sigma_T \nu n_{\rm e}}{c}\bigg)^{1/3} \approx 5 \: \rho_{7}^{2/3} \, \nu_{9}^{1/3} \, n_{{\rm e},12}^{1/3}
\end{split}
\end{equation}
Where we have used Eq. \ref{eq:B_perp}. We find that synchrotron radiation is almost always subdominant, and does not constrain coherent curvature radiation. The small scale gyration leads to simultaneous high-energy radiation, especially outside of the coherent region where the gyration will dominate particle cooling. We estimate and discuss such emission in Section \ref{sect:highenergy}.

\subsection{Constraint due to momentum misalignment}
\label{sect:momenta_cond}
For the radiation to be coherent, given the constraint in Eq. \ref{eq:B_constraint}, we require that:
\begin{equation}
\begin{split}
        B &> \gamma B_{\perp} 
        = 8 \pi e n_{\rm e} \gamma^2 c \nu^{-1} \\
        \gamma &< \bigg(\frac{B \nu}{8 \pi e n_{\rm e} c}\bigg)^{1/2} \\
        \gamma_{\rm max} &\approx  500 \: B_{11}^{1/2} \, \nu_{9}^{1/2}\, n_{{\rm e},12}^{-1/2} \\
\end{split}
\label{eq:lorentz}
\end{equation}
Here we have used Eqs. \ref{eq:B_perp} and typical magnetized neutron star (NS) parameters, and find that particle bunches with large lorentz factors induce a magnetic field which destroys coherence. In most situations from which we expect coherent radiation, it is thought the number density of particles scales with the magnetic field $B$ \citep{GoldreichJulian1969} as approximately:
\begin{equation}
\begin{split}
        n_{\rm e} &= \xi n_{GJ} = \dfrac{2 \xi B_s R_{\rm NS}^3}{e c P R^3} = 1.4 \times 10^{12} \; B_{s,11} \,P_{-1}^{1/2}\, \xi_{1}\, R_{6}^{-3} \,{\rm cm^{-3}}
\end{split}
\label{eq:goldreich}
\end{equation}
Where $P$ is the NS period, $R_{\rm NS} = 10^{6}$ cm is the NS radius, $R \geq R_{\rm NS}$ is the distance from the NS centre, $B_s$ is the dipole surface magnetic field and $\xi > 1$ is the pair multiplicity due to photon-magnetic field interactions producing pairs. We assume the leptons originate from pair creation, so there is charge neutrality. We can rewrite Eq. \ref{eq:lorentz} explicitly in terms of the NS parameters:
\begin{equation}
    \begin{split}
        \gamma_{\rm max} &= \bigg(\frac{P \nu}{16 \pi \xi}\bigg)^{1/2} = 500 \: P_{-1}^{1/2} \, \nu_{9}^{1/2}\, \xi_{1}^{-1/2}
    \end{split}
    \label{eq:lorentz_goldreich}
\end{equation}
To obey Eq. \ref{eq:deltamomenta}, we should also require that all field lines occupied by the coherent patch be well aligned. Assuming a dipole field, and that the transverse source size extends from $R$ above the polar cap to $(R,\delta \theta)$, we find that this could further limit emission close to the NS surface:
\begin{equation}
    \begin{split}
        \frac{1}{\gamma} &> \frac{\sin(\delta \theta) R_{\rm NS}^3}{R^3} 
        \approx \frac{R_{\rm coh} R_{\rm NS}^3}{R^4} \\
        \gamma &< \bigg(\frac{R^4 \nu}{R_{\rm NS}^3 c}\bigg)^{1/2} \approx 180 \: R_6^{2} \, \nu_9^{1/2}  \\
    \end{split}
    \label{eq:fieldlines}
\end{equation} 
However, a source with a transverse size less than $R_{\rm coh}$ can have higher lorentz factors. 

\subsection{Constraints on duration}
\label{sect:duration}
If the decay timescale of the accelerating electric field is large, we expect many patches of coherent emission $N_{\rm p,l}$ to extend along the observer's line of sight. The observed duration of coherent curvature radiation is then limited by either the observer frame light crossing time of the patches: $\frac{N_{\rm p,l} R_{\rm coh}}{\gamma c} = N_{\rm p,l}/\nu$, the sweep of the radiation beam: $\frac{N_{\rm p,l} \rho}{\gamma c}$, or the movement of particles along field lines into regions of lower field strength such that coherence cannot be supported via Eq. \ref{eq:lorentz}. In all cases, bursts that originate closer to the NS surface are expected to be shorter in duration due to smaller spatial scales, tighter field lines \citep{Bilous2019} and more rapidly decreasing magnetic field strength.

\section{Predictions}
\label{sect:Predictions}

\subsection{Maximum luminosity of coherent curvature radiation}
\label{sect:max_lum_coh}
\begin{figure}
    \centering
    \includegraphics[width=.5\textwidth]{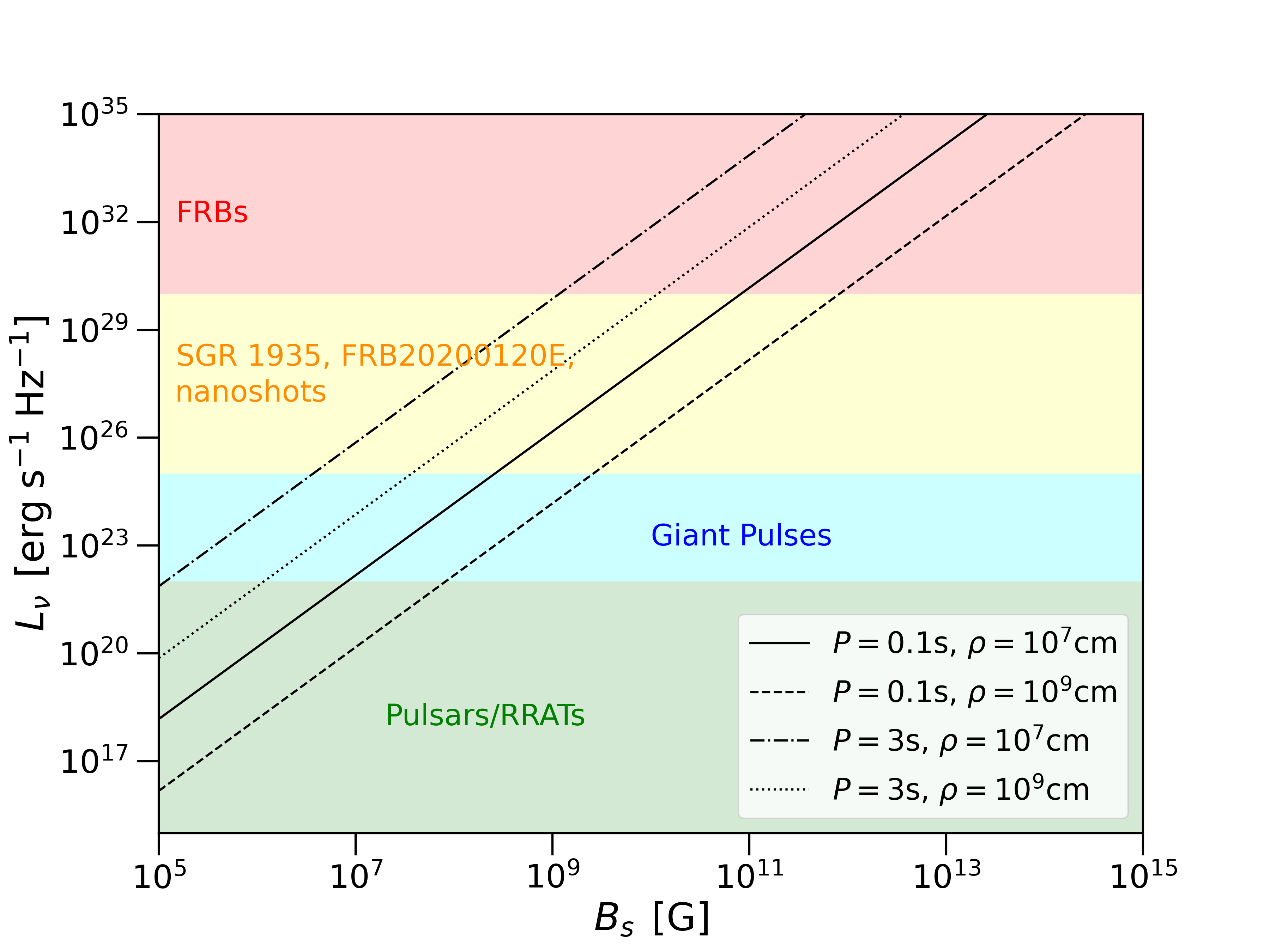}
    \caption{We plot the maximum spectral luminosity for two reasonable limiting source parameters using Eq \ref{eq:Lnu_pulsar}. We fix $\xi = 10$, $N_{\rm p} = 1$, $\nu = 10^9$ and $R = 2 \times 10^6$ cm. In the background we show typical spectral luminosities of coherent radio sources \citep{Pietka2015}, noting in particular the sources in yellow that bridge the gap between extra-galactic and galactic sources \citep{Hankins2007,Bochenek2020,Nimmo2021}.}
    \label{fig:Simple_lum_plot}
\end{figure}

Given the condition in Eq. \ref{eq:lorentz_goldreich}, we can derive a maximum emitted luminosity of coherent curvature radiation given source parameters:
\begin{equation}
    \begin{split}
        L_{\rm coh, max} &= \frac{2 N_{\rm p}(N e)^2 c \gamma_{\rm max}^4}{3 \rho^2} = N_{\rm p} n_{\rm e}^2 R_{\rm coh}^6 \frac{2 e^2 c \gamma_{\rm max}^4}{3 \rho^2} \\
        &= 2 \times 10^{37} \: B_{s,11}^{2} \, P_{-1}^{3} \,\rho_{7}^{-2} \,\nu_{9}^{-1} \,\xi_{1}^{-3} \, N_{\rm p} \, R_{6}^{-6} \: {\rm erg \, s^{-1}}
    \end{split}
    \label{eq:max_coh_lum}
\end{equation}
Where we have used Eqs. \ref{eq:R_coh} and \ref{eq:lorentz_goldreich}, $\rho$ is the magnetic field line curvature radius and $N_{\rm p}$ is the number of coherent patches that add to the luminosity incoherently. The maximum observed spectral luminosity is approximately $L_{\nu, \rm obs} = \gamma^2 L/\nu_c$ where $\nu_c = 3 c \gamma^3/4 \pi \rho$ and the $\gamma^2$ factor is due to beaming of emission into a small observable solid angle \citep{Lyutikov2021}: 
\begin{equation}
\begin{split}
    L_{\nu, \rm max}^{\rm obs} &= 6 \times 10^{31}  \:  B_{s,11}^{2} \, P_{-1}^{5/2}\, \rho_{7}^{-1} \, \nu_{9}^{-3/2} \,\xi_{1}^{-5/2} \, N_{\rm p} \, R_{6}^{-6} \; {\rm erg \, s^{-1} \, Hz^{-1}}\\
    T_{\rm B, max}^{\rm obs}  &= 
    \frac{2 c^2 L_{\nu}^{\rm obs}}{k_B \nu^2 R_{\rm coh}^2}\\ 
    &= 4 \times 10^{42} \: B_{s,11}^{2} \, P_{-1}^{3/2} \,\rho_{7}^{-1} \,\nu_{9}^{-5/2} \,\xi_{1}^{-3/2} \, N_{\rm p}\, R_{6}^{-6} \; {\rm K}
\end{split}
\label{eq:Lnu_pulsar}
\end{equation}
This upper limit to the spectral luminosity fits well with the observed maximum spectral luminosity from extreme galactic coherent sources as shown in Fig. \ref{fig:Simple_lum_plot}. Except for FRBs, these pulses represent the brightest coherent radio emission observed, suggesting a common coherent curvature mechanism for giant pulses and FRBs \citep{Keane2012,Cordes2016}. Eq. \ref{eq:Lnu_pulsar} refers to the $\gamma = \gamma_{\rm max}$ maximal case, for non-maximal bursts with $\gamma < \gamma_{\rm max}$ the luminosity drops rapidly: $L_{\nu}^{\rm obs} \propto R_{\rm coh}^6 \gamma^3 \propto \gamma^{9}$. 


\subsection{Coincident incoherent high-energy emission}
\label{sect:highenergy}
There is growing evidence that the mechanism responsible for coherent radio emission is also powers emission at higher energies \citep{Younes2021,Enoto2021,HAWC2021}. In Section \ref{sect:gyrationcooling} we discussed the possibility of a subdominant radiation mechanism due to small-scale gyrations caused by the induced field. We can look at the power and critical frequency of the emission, assuming it manifests as incoherent synchrotron radiation about $B$ with an angle $\alpha = B_{\perp}/B$ as discussed in Section \ref{sect:particle_motion}:
\begin{equation}
    \begin{split}
        L_{\rm sync} &= N_{\rm p} n_{\rm e} P_{\rm sync}
        = 16 \pi \zeta^5 c^6 \sigma_T n_{\rm e}^3 e^2 N_{\rm p} \gamma_{\rm max}^7 \nu^{-5} \\
        &= 5 \times 10^{31}  \: B_{s,11}^{3} \, P_{-1}^{1/2} \,\nu_{9}^{-3/2}\,\xi_{1}^{-1/2} \, \zeta^{5} \,  N_{\rm p}\,  R_{6}^{-9} \; {\rm erg \, s^{-1}}
    \end{split}
    \label{eq:L_sync}
\end{equation}
The total transverse particle acceleration region may be larger than $R_{\rm coh}$ by a factor of $\zeta$, as long as the total size does not exceed $\approx R_{\rm NS}$. Accelerated particles outside of the coherent region will follow helical field lines and radiate incoherently,
so the radio is much suppressed relative to the high-energy emission. The larger transverse size means these particles have a larger pitch angle and a large emission volume such that the incoherent luminosity scales as $\zeta^{5}$. We have assumed particles outside of the incoherent region will have approximately the same lorentz factor as the coherent particles, which may not be the case. Furthermore, we note sometimes only field lines outside of the coherent region point towards the observer such that only the high-energy radiation is visible. The critical frequency of this emission is:
\begin{equation}
\begin{split}
    \nu_{\rm c,sync} &= \frac{3}{2} \gamma_{\rm max}^3 \omega_B \sin(\alpha) \\
    &\approx 10^{21}  \: P_{-1}^{1/2} \, \nu_9^{1/2} \, B_{s,11} \, \xi_1^{-1/2} \, \zeta \, R_6^{-3} \; {\rm Hz}
    \end{split} 
    \label{eq:crit_freq_sync}
\end{equation}
Or approximately $E_{\rm ph} = 5 \, {\rm MeV}$. We note that the observed cut-off will be below this critical frequency in high-field sources, due to photo-magnetic processes of photon splitting and pair production \citep{DaughertyHarding83}. Observations of a high-energy cut-off are usually dominated by one photon pair production, and could be used as a diagnostic of the local magnetic field $B$. Assuming $\mathcal X \ll 1$, which holds for the low-energy cut-off in non-critical fields, the photon attenuation factor is approximately \citep{HardingBaring1997}: 
\begin{equation}
    T_{\rm pp} \approx \frac{0.3 \alpha m_{\rm e} c}{\hbar} \frac{B}{B_c} \exp\bigg(\frac{-4}{3 \mathcal X} \bigg) \:  \: \: \: \: {\rm where} \: \: \mathcal X = \frac{E_{\rm ph}}{2 m_{\rm e} c^2} \frac{B}{B_c}
\end{equation}
Where we have made simplifying assumptions that photons propagate a distance comparable to the curvature radius $\rho$ such that $\sin(\theta_{\rm k B}) \approx 1$, and that the drop in $B$ is negligible across this distance. If photons are attenuated if $T_{\rm pp} > 1$, we find an approximate maximum energy cut-off of a few MeV for $B = 10^{13}$ G and around 10 GeV for $B = 10^{9}$ G. The emission spectrum of this incoherent emission is expected to follow a synchrotron spectrum and thus for $\nu < \nu_c$, we estimate the observed spectral luminosity as:
\begin{equation}
    \begin{split}
        &L_{\nu, \rm sync}^{\rm obs} = \frac{4 \gamma^2 L_{\rm sync}}{3 \nu_{\rm c,sync}} \bigg(\frac{\nu_{\rm x}}{\nu_{\rm c,sync}}\bigg)^{1/3} \\
        &= 8 \times 10^{14} \; B_{s,11}^{5/3}  \,P_{-1}^{5/6} \nu_{\rm x,18}^{1/3} \,\xi_{1}^{-5/6} \,\nu_{9}^{-7/6} \,\zeta^{11/3} \,N_{\rm p} \,R_{6}^{-5} \; {\rm erg \, s^{-1} \, Hz^{-1}}
    \end{split}
    \label{eq:L_nu_sync}
\end{equation}
Using Eqs. \ref{eq:Lnu_pulsar} \& \ref{eq:L_nu_sync} we can estimate the ratio of radio/X-ray flux in representative bands by assuming luminosity across a bandwidth $\delta \nu$ centred on $\nu$ is approximately $L_{\nu} \delta \nu$, where $\delta \nu \approx \nu$:
\begin{equation}
\begin{split}
    \frac{F_{0.1-1 \rm GHz}}{F_{1-10 \rm keV}} & \approx 7 \times 10^{7} \: B_{s,11}^{1/3}  \,P_{-1}^{5/3}\, \rho_{7}^{-1} \, \nu_{\rm x,18}^{-4/3} \,\xi_{1}^{-5/3} \,\nu_{9}^{2/3} \, \zeta^{-11/3}\,R_{6}^{-1}
\end{split}
\label{eq:flux_ratio}
\end{equation}

\subsection{Crab Pulsar}
The Crab Pulsar produces kilo-Jansky flux giant pulses at GHz frequencies \citep{Lundgren1995}, which represents a spectral luminosity of approximately $L_{\nu} \approx 5 \times 10^{24} \; {\rm erg s^{-1} \, Hz^{-1}}$, thought to come from high altitudes close to the light cylinder \citep{EilekHankins2016}. We can place limits on emission parameters using source parameters for the Crab ($B_s = 7 \times 10^{12} $G, $P = 0.033$s; \citealt{Lyne1993}) and Eq. \ref{eq:Lnu_pulsar} to solve for $R$, assuming coherent curvature radiation. We further assume a pure dipole magnetic field, $\rho = 10^7$ and $\xi = 10$. We find the origin of a kJy burst must be less than 400km from the surface of the NS. The inferred limits on parameters of the emission are: $B \geq 10^{8}$G and $20 < \gamma \leq 250$ depending on $\frac{R}{R_{\rm NS}}$, but emission closer to the NS with modest lorentz factors is preferred due to causality arguments in Section \ref{sect:Induced magnetic field}.
\par
The most extreme Crab nanoshot had a 9 GHz flux of 2 Mega-Jansky ($L_{\nu} \approx 10^{28} \; {\rm erg \, s^{-1} \, Hz^{-1}}$; \citealt{Hankins2007}). Again via Eq. \ref{eq:Lnu_pulsar}, we find that these brightest nanoshots must originate from less than 60km from the neutron star surface, assuming fiducial parameters. Despite the uncertainties involved in estimates of $B_s$ and $\xi$, the dependence of $R^{-6}$ in Eq. \ref{eq:Lnu_pulsar} means this result is very constraining even for large uncertainties in the source parameters. The short duration of these bursts is also consistent with discussion in Section \ref{sect:duration} given how close to the NS the emission originates.
\par
Recently \cite{Enoto2021} observed for the first time a $3\%$ increase in the 0.2-12 keV X-ray flux associated with Crab giant pulses ($\nu=2$ GHz), detecting a flux increase of $\delta F \approx 8  \times 10^{-10} \; {\rm erg \, s^{-1} \, cm^{-2}}$. We can estimate the 0.2-12 keV flux associated with the brightest giant pulses ($\approx 10 $kJy) observed by \cite{Enoto2021} via Eq. \ref{eq:flux_ratio}, assuming $B_s$ and $P$ as before. We find that fiducial parameters can explain both fluxes simultaneously at a distance for maximal bursts originating $\approx 200$ km from the NS surface if $\zeta \approx 30$. The implied coherent and incoherent emission regions have transverse sizes of $5 \times 10^{3}$ cm and $1.5 \times 10^5$ cm respectively. We therefore suggest that small-scale particle gyration due to the induced field could plausibly explain the X-ray flux observed by \cite{Enoto2021}, and the X-ray/radio flux ratio could be used to constrain the location of giant pulse emission. Non-detections of higher energy emission by other observatories, particularly the stringent upper limit reported in \cite{Magic2020}, is line with predictions of emission from close to the NS surface as higher energy photons are attenuated as discussed in Sect. \ref{sect:highenergy}. We note that in \cite{Enoto2021}, the authors discuss possible origins of the increased X-ray flux during giant pulses which are not related to the coherently emitting particles themselves.  
\subsection{SGR 1935+2154} 
On 27th April 2020 a bright radio burst was observed from SGR 1935 with a $1.4$ GHz spectral luminosity of $L_{\nu} = 1.6 \times 10^{26} \, {\rm erg \, s^{-1} \, Hz^{-1}}$ \citep{Bochenek2020}. Assuming $B_s = 2.2 \times 10^{14}$ G and $P = 3.24$ s \citep{Younes2021}, Eq. \ref{eq:Lnu_pulsar} suggests the maximum distance of approximately 4000 km from the magnetar's surface assuming fiducial parameters. Furthermore, a coincident X-ray burst was observed with an 100 keV luminosity of approximately $10^{39} \, {\rm erg \, s^{-1}}$ \citep{Mereghetti2020,Li2021,Ridnaia2021}, with a harder spectrum than other magnetar bursts from the source \citep{Younes2021}. We find that both fluxes can be explained simultaneously for a maximal burst only if we allow non-fiducial parameters e.g. $\rho \approx 10^{10}$cm and $\zeta \approx 100$. Nevertheless, the prediction of high-energy emission with the same beaming factor as the coherent emission can explain the peculiar spectra of the X-ray burst temporally coincident with the radio burst. We suggest coherent curvature radiation could be a universal feature of magnetar X-ray bursts, but observable only for a small fraction of cases due to the beaming restriction, whereas the thermal quasi-isotropic X-ray emission is observed more often.

\subsection{Caveats and FRBs}
There are a few caveats to the luminosity upper limit in Eq. \ref{eq:max_coh_lum}. Firstly, we have assumed an approximately spherical source. A coherent source with longitudinal size $R_{\rm l}$ and transverse size $R_{\rm t}$ where $R_{\rm coh } > R_{\rm l} > R_{\rm t}$ would induce a smaller current and therefore allow luminosities up to a factor of $R_{\rm l}/R_{\rm t}$ larger. Furthermore, we have not considered in detail sources with $N_{\rm p} > 1$, where the number and geometry of the patches affects both the luminosity and duration of the observed radiation. We also note it is possible that some essential property of coherent curvature radiation prevents the emission of simultaneous high-energy radiation as predicted in Section \ref{sect:highenergy}. 
\par
The limitations described here still apply in quantum critical magnetic fields as was outlined in \cite{Kumar2017}, where the authors suggest that FRBs must occur very close to the surface of magnetars in fields $B > 10^{14}$G. If particles confined to the ground state follow the total $B + B_{\perp}$ field, they should still gyrate as described in Section \ref{sect:particle_motion} and thus we might expect high-energy emission. Using Eq. \ref{eq:L_sync} we predict X-ray emission of $\approx 10^{44}\, {\rm erg s^{-1}}$ below $1$\,MeV to accompany maximal magnetar ($B_s = 10^{15}\,{\rm G}$) bursts. We can compare our prediction to the X-ray limits of \cite{Scholz2017} using Eq. \ref{eq:flux_ratio} and assuming a 1 millisecond burst duration. We find a $0.5-10$\,keV fluence of approximately $10^{-26} \: {\rm erg \, cm^{-2}}$ to accompany radio bursts of $0.5$ Jy, well below the constraints. We note that isotropic magnetar burst emission may dominate, depending on the parameters.

\section{Conclusion}
\label{sect:Conclusion}
In this Letter we have considered electrodynamic interactions between coherently radiating particles. We have shown in Section \ref{sect:Limitations} \& \ref{sect:Predictions} that there is an upper limit to the radio luminosity of coherent curvature radiation which depends on the source parameters. This limit suggests that if the giant pulses are powered by coherent curvature radiation, they must originate in the inner magnetosphere very close to the NS surface. Furthermore, small scale particle gyration could mean that coherent curvature radio pulses are universally associated with high-energy counterparts. A common coherent curvature radiation origin of giant pulses and FRBs can be falsified by observations of emission from a known source more luminous than allowed by the limits in Fig. \ref{fig:Simple_lum_plot}. Future work includes investigating the quantitative effect of multiple coherent patches, the frequency and polarization predictions of coherent curvature emission taking into account individual particle gyration on small scales and modelling giant pulse \& FRB populations.  

\section*{Acknowledgements}
We would like to thank J. I. Katz \& M. Lyutikov for valuable discussion, and the referee P. Kumar for helpful comments which improved this work. AC is supported by the Netherlands Research School for Astronomy (NOVA). 

\section*{Data Availability}
A Python notebook from which the results and figures of this Letter can be reproduced will be made available at DOI:10.5281/zenodo.5211149


\bibliographystyle{mnras}
\bibliography{references} 





\bsp	
\label{lastpage}
\end{document}